\documentclass[sigconf,nonacm]{acmart}

\AtBeginDocument{%
  \providecommand\BibTeX{{\normalfont B\kern-0.5em{\scshape i\kern-0.25em b}\kern-0.8em\TeX}}}

\usepackage{booktabs}
\usepackage{graphicx}
\usepackage{amsmath}

\renewcommand\footnotetextcopyrightpermission[1]{}

\begin{document}

\title[Multimodal Asset Personalization]{Multimedia Asset Personalization
  via Multimodal Embeddings at Netflix}

\author{Emma Yanyang Kong}
\affiliation{%
  \institution{Netflix}
  \city{Los Gatos}
  \state{California}
  \country{USA}}
\email{ekong@netflix.com}

\author{Aditya Deshpande}
\affiliation{%
  \institution{Netflix}
  \city{Los Gatos}
  \state{California}
  \country{USA}}
\email{adityad@netflix.com}

\author{Bowei Yan}
\authornote{Work done while at Netflix.}
\affiliation{%
  \institution{}
  \city{}
  \state{}
  \country{}}
\email{yanbowei@gmail.com}

\author{Asad Abbasi}
\affiliation{%
  \institution{Netflix}
  \city{Los Gatos}
  \state{California}
  \country{USA}}
\email{aabbasi@netflix.com}

\author{Santiago Castro}
\affiliation{%
  \institution{Netflix}
  \city{Los Gatos}
  \state{California}
  \country{USA}}
\email{santiagoc@netflix.com}

\author{Avneesh Saluja}
\authornotemark[1]
\affiliation{%
  \institution{Cohere}
  \city{San Francisco}
  \state{California}
  \country{USA}}
\email{asaluja@gmail.com}

\author{David Fagnan}
\affiliation{%
  \institution{Netflix}
  \city{Los Gatos}
  \state{California}
  \country{USA}}
\email{dfagnan@netflix.com}

\author{Ashish Rastogi}
\affiliation{%
  \institution{Netflix}
  \city{Los Gatos}
  \state{California}
  \country{USA}}
\email{arastogi@netflix.com}

\renewcommand{\shortauthors}{Kong et al.}

\begin{abstract}
Personalized promotional assets, namely artwork images and video preview clips,
are critical to content discovery on Netflix. Traditional recommender models for
asset selection rely on members' interaction history with IDs, leaving
them blind to asset content and challenging to serve newly launched titles and
their associated assets.
We describe how multimodal embeddings have reshaped production systems
and report transferable lessons for practitioners adopting
foundation-model embeddings into recommender systems.
\emph{First, pretrained image embeddings unlock cross-title, cross-canvas knowledge
transfer.} Augmenting a two-tower model with CLIP~\cite{radford2021clip}
image embeddings allows a single model to serve all five Netflix artwork canvas
types used across different devices, replacing five separately trained
per-canvas models and substantially improving performance on cold-start items. A
lightweight extension reuses CLIP's joint text-image space to make
artwork personalization query-aware in the search context.
\emph{Second, multimodality decisively beats any single modality for video
preview personalization.} We describe \emph{MediaFM}, our in-house tri-modal
foundation model trained on a large-scale corpus of shots from the Netflix show catalog,
fusing visual (\emph{SeqCLIP}), audio (wav2vec~2.0~\cite{baevski2020wav2vec}), and
timed-text signals, and adopt it for video preview personalization; MediaFM
outperforms strong visual-only baselines both offline and in online A/B tests
across platforms.
\emph{Third, a simple offline proxy task whose performance correlates with online
outcomes can accelerate the experimentation and productization cycle.}
Predicting the popularity-based winner from embeddings
alone ranks embedding models and versions, letting us prune the choice space \emph{before} any
end-to-end model integration and online A/B testing, and now gates
every new MediaFM checkpoint before release.
In this paper, we also share our production engineering decisions (shared
embedding infrastructure, low-latency serving, and cheap screening) that made
these deployments viable, and the design tradeoffs and some failure modes we
encountered along the way.
\end{abstract}



\maketitle

\pagestyle{plain}
\thispagestyle{plain}

\section{Introduction}
Every Netflix title is represented to members through promotional assets:
artwork images shown as tiles across multiple canvases, and short video
previews that autoplay during browsing.
Netflix selects the specific image or video preview for each member
algorithmically~\cite{chandrashekar2017artwork}, but traditional models for
this task are \emph{content-blind}. They treat each asset as an opaque ID
and learn nothing transferable about its visual or semantic content.
This creates a persistent \emph{cold-start} problem: newly launched titles and
freshly created assets have very little interaction history, only a small
amount gathered during an initial exploration phase, so the model falls back to
popularity heuristics that ignore individual member preferences.

Formally, for each (member $u$, title $t$) pair, the system selects an
asset $a_t^\star(u) \in A_t$ from a candidate set
$A_t = \{a_{t1}, \dots, a_{tn_t}\}$ so as to maximize a member interaction
reward $r(u, a_{ti})$. Traditional models estimate $r$ from logged
$(u, a_{ti})$ interactions alone, which makes them blind to asset
content and breaks down whenever $A_t$ contains newly launched assets
with no logged history.

We address this by integrating multimodal foundation model embeddings into our
personalization models, giving them genuine content awareness.
This paper describes two production systems deployed at Netflix:

\begin{enumerate}
\item \textbf{CLIP-Augmented Artwork Personalization} (\S\ref{sec:artwork}):
  We concatenate CLIP image embeddings~\cite{radford2021clip} into the item
  tower of a two-tower retrieval model, enabling cross-title and
  cross-canvas knowledge transfer and collapsing five per-canvas
  models into \textbf{a single unified architecture} that transfers
  signal from high-impression canvases, titles, and assets to
  low-impression ones.

\item \textbf{Tri-Modal Video Preview Personalization} (\S\ref{sec:video}):
  We adopt \emph{MediaFM}, an in-house multimodal foundation model that fuses
  visual, audio, and timed-text signals, trained on a large-scale corpus of shots from the Netflix show catalog. 
  As a secondary contribution, we provide a detailed description of MediaFM's architecture, pretraining, and
  intrinsic evaluation (Appendix~\ref{app:mediafm}), previously documented only
  in the company's technology blog.
\end{enumerate}

Both systems use the Netflix Embedding Store~(\S\ref{sec:overview}) for
embedding retrieval at training and serving time, and both are validated by
IPS-weighted offline metrics~(\S\ref{sec:ips}) and large-scale online A/B tests.
A separate methodological contribution (\S\ref{sec:proxy}) is a cheap,
offline proxy task (popularity-based winner prediction over asset
candidates per title) that screens embedding candidates before any
end-to-end offline experiments or online A/B traffic are allocated.

Our contribution is not a new
learning algorithm (concatenating pretrained embeddings into a two-tower
model is deliberately standard) but the \emph{production integration}: how a
shared embedding platform, a low-latency serving path, and a cheap offline
screening gate together turn off-the-shelf foundation-model embeddings into
shipped, measurable member-facing wins. Throughout, we report our design
decisions, tradeoffs, and failure modes (\S\ref{sec:artwork-ablation},
\S\ref{sec:proxy-justification}, \S\ref{sec:lessons}) that a practitioner would
need to reproduce the outcome in a different organization.

\section{Related Work}
\label{sec:related}

\paragraph{Two-tower retrieval.}
Two-tower models, which embed two sides of a matching problem into a shared
space where relevance is an inner product, originate in web search with
DSSM~\cite{huang2013dssm} and have become the workhorse of industrial retrieval
and candidate generation~\cite{covington2016deep}. We adopt the standard
\emph{two-tower} form in which a dedicated item tower encodes item content
features alongside identifiers~\cite{yi2019sampling}. Such models scale to large
corpora because item representations do not depend on the user and can therefore
be computed ahead of time and served by approximate nearest-neighbor
search~\cite{johnson2021faiss}, a property we lean on heavily
(\S\ref{sec:serving}). Our systems keep this architecture unchanged and
intervene only in the item tower, which is what makes the integration cheap:
the retrieval machinery, training pipeline, and serving stack are untouched.

\paragraph{Content and multimodal signals in recommendation.}
A long line of work incorporates content features to combat cold-start and
enrich item representations, from visually-aware collaborative
filtering~\cite{he2016ups} to graph- and attention-based fusion of multimodal
side information~\cite{wei2019mmgcn,liu2021noninvasive}; see~\cite{liu2024mmrecsurvey}
for a recent survey of the multimodal recommendation landscape. Our contribution is
orthogonal to the specific fusion mechanism: we treat the foundation-model
embedding as a frozen, platform-served feature and study what it takes to adopt
it across several production systems, rather than proposing a new fusion
architecture.

\paragraph{Foundation-model embeddings.}
CLIP~\cite{radford2021clip} provides a joint image--text space that we exploit
both for content-aware artwork ranking and for query-aware ranking
in search. For video previews, we build on self-supervised audio
representations~\cite{baevski2020wav2vec} and large text embedding
models~\cite{openai2024embeddings}, fused by our in-house MediaFM
model~\cite{saluja2026mediafm}; video--language understanding has an
established benchmark literature~\cite{xu2016msrvtt} and a growing family of
self-supervised video foundation models based on masked
modeling~\cite{tong2022videomae} and video--language
pretraining~\cite{wang2022internvideo} that motivate tri-modal
representations for promotional clips. Efforts such as
ImageBind~\cite{girdhar2023imagebind} further point toward a single embedding
space spanning many modalities, which informs our forward-looking goal
(\S\ref{sec:conclusion}).

\paragraph{Counterfactual (off-policy) evaluation.}
Inverse-propensity scoring~\cite{schnabel2016recommendations,saito2021counterfactual}
lets us compare candidate policies offline on logged data. Unlike the common
setting where propensities are \emph{estimated}, we log them exactly from a
randomized exploration policy (\S\ref{sec:ips}), which removes a major source
of bias and is, in our experience, the single most important ingredient for
offline metrics that track online outcomes.

\paragraph{Asset personalization at Netflix.}
Artwork personalization at Netflix was first described
in~\cite{chandrashekar2017artwork} as a contextual-bandit problem over a fixed
per-title artwork inventory; related work on automatic thumbnail
selection~\cite{song2016thumbnails} scores individual frames but is not
personalized. We extend that line of work in three ways: content
awareness via foundation-model embeddings, consolidation of per-canvas models
into one, and an extension of the same embeddings to video previews and to
query-aware ranking in search.

\section{System Overview}
\label{sec:overview}

All systems in this paper share a common backbone: the \emph{Netflix
Embedding Store}, a component of Netflix's AI Platform that hosts dense
embeddings for titles, games, and member profiles, and serves them with
training-time and online-inference parity.
Its key property for our work is that it \emph{decouples foundation-model
updates from personalization-model deployments}: new embeddings (or new
versions of existing ones) can be registered, backfilled, and validated independently
without coordinating downstream application models.
As a result, a new foundation model becomes available to every ranking system or
personalization model through registration and configuration alone.
Versioning is explicit: a new embedding version is written under a new key and
backfilled offline, so a downstream model can adopt it by configuration and roll
back instantly if a regression is detected.

\begin{figure}[htbp]
  \centering
  \includegraphics[width=\linewidth]{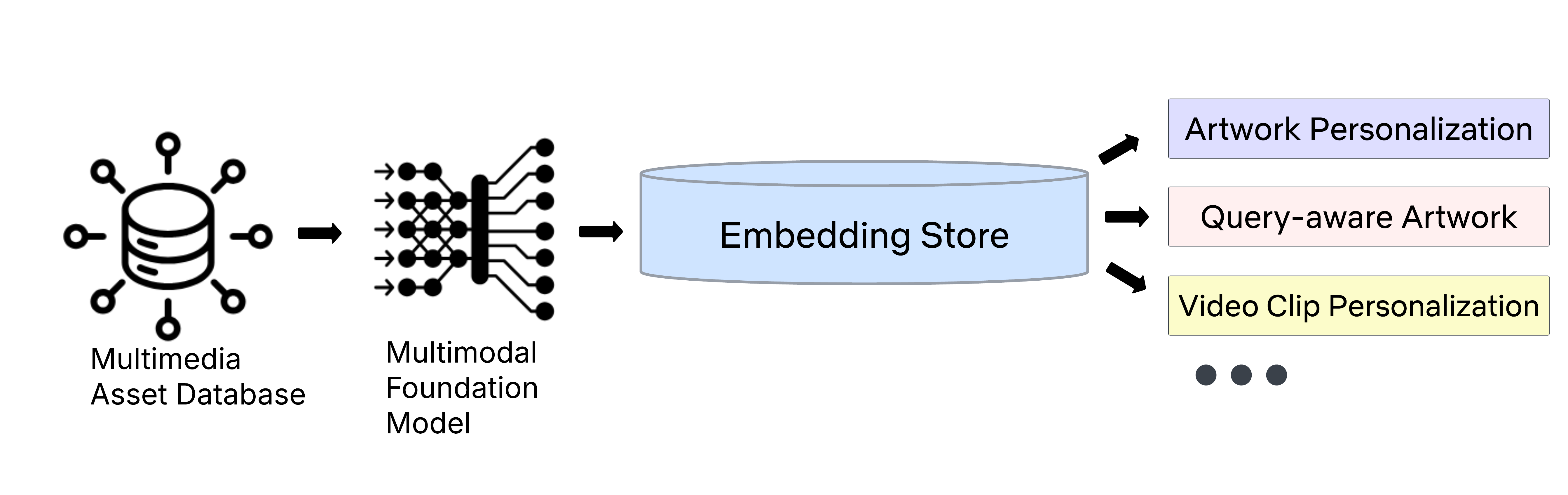}
  \caption{Shared embedding infrastructure. Foundation model embeddings are stored
    in the Netflix Embedding Store and consumed by downstream personalization systems:
    artwork personalization, query-aware artwork, video preview personalization, other rankers, etc.}
  \Description{Block diagram showing foundation models (CLIP, SeqCLIP, MediaFM) feeding
    embeddings into the Netflix Embedding Store, which in turn feeds three
    two-tower personalization models for artwork, query-aware artwork, and video.}
  \label{fig:overview}
\end{figure}

Figure~\ref{fig:overview} illustrates the shared architecture.
Foundation models (CLIP, SeqCLIP, MediaFM, etc.) encode raw asset
contents into dense vectors that are stored in the Embedding Store and
consumed at both training and inference time by the item towers
of the two-tower retrieval models described in the following sections.

This decoupling design enables new foundation models to be integrated into any personalization
system by registering their embeddings in the Embedding Store, without modifying
the downstream training or serving infrastructure.

\subsection{Production Serving and Cost}
\label{sec:serving}

A recurring practical concern when moving from ID-only models to
multimodal-augmented item towers is whether the heavier representations can be
served within the strict latency budget of homepage and search page
generation. In our deployment the question largely dissolves, because nothing on
the multimodal path executes during page generation: the system is
\emph{precompute-first}.

\paragraph{Three layers of precomputation.}
Neither an asset's content nor its score depends on the incoming request, which
lets us push work to precomputation at three levels.
\emph{(i) Content embeddings.} A title's artwork and previews do not change
between requests, so their foundation-model embeddings are computed once at
asset-ingestion time and written to the Embedding Store; CLIP and MediaFM are
never invoked online.
\emph{(ii) The item-tower projection.} The item tower's output $\mathbf{h}_a$
(Eq.~\ref{eq:clip-item}), which fuses an asset's ID and content embeddings,
depends only on the asset and is therefore identical across all member profiles,
so we compute and cache it once instead of re-running the item tower per
request.
\emph{(iii) The selection itself.} The asset personalization models compute
the selected asset at the (member profile, show title) level on a daily cadence.
A page-generation request therefore reduces to a key-value lookup of an
already-materialized selection. This is why the heavier multimodal item tower
meets the same latency budget as the ID-only baseline, and why adopting CLIP and
MediaFM required no relaxation of production latency SLAs: the request path is
structurally identical under both, and the two differ only in what was written
during the precomputation phase.

\paragraph{Precomputing only each profile's top titles.}
Layer (iii) is deliberately \emph{not} applied to the full catalog for every
profile. We materialize selections only for each profile's top ${\sim}500$
titles, those most likely to be surfaced to them, and elsewhere fall back to the
unpersonalized popularity-based winner (\S\ref{sec:proxy-winner}), the asset that
performs best under a uniform serving policy. Because a member's impressions
concentrate on their most relevant titles, this per-profile head covers the large
majority of what they actually see while cutting precomputation cost by up to
$50\%$, a deliberate cost--coverage tradeoff that keeps the system economical at
scale without materially affecting the member experience.

\section{Offline Evaluation with IPS}
\label{sec:ips}
Throughout this paper, we evaluate models offline through Inverse Propensity
Score (IPS) weighting~\cite{schnabel2016recommendations}.
Evaluating on logs from the production policy can be biased. The
production policy chooses some assets far more often than others, so
observed rewards reflect what the \emph{policy} preferred, not what the
member would have preferred under counterfactual choices.
A new model's performance on such logs is therefore systematically
underestimated whenever it disagrees with the logging policy.
To avoid this bias and to obtain propensities that are \emph{known by
construction} rather than estimated, we evaluate on the dedicated
\emph{explore data}: a small fraction of member traffic is held out and
served by a randomized exploration policy that samples among candidate
assets according to a known distribution.
Because we control the exploration policy, the propensity
$p(a \mid x)$ of each impression, for action $a$ given context $x$, is logged exactly at serving time, with
no model-based estimation step and no positivity assumptions to defend.
IPS reweights each explore-data observation by the inverse of this
logged propensity:
\begin{equation}
  \hat{R}_{\text{IPS}} = \frac{1}{|D|}
    \sum_{(x, a) \in D} \frac{r(x, a)}{p(a \mid x)}
  \label{eq:ips}
\end{equation}
where $D$ is the explore-data slice, $r(x, a)$ is the
observed reward (e.g., play), and $p(a \mid x)$ is the recorded
exploration propensity. Under these conditions
$\hat{R}_{\text{IPS}}$ is an unbiased estimate of the reward a new
policy would have obtained had it been deployed, which lets us compare
new and production models on the same data without running an
A/B test.
We use this estimator as the primary offline metric for both artwork
(\S\ref{sec:artwork}) and video preview (\S\ref{sec:video}) personalization.
We report all results as ratios relative to the production baseline.

\section{Artwork Personalization}
\label{sec:artwork}

\subsection{Cold-Start, Knowledge Transfer, and Model Consolidation with CLIP}
\label{sec:artwork-clip}

For each Netflix title, we maintain a diverse inventory of static artworks
across multiple display canvases and devices: billboard, vertical box, horizontal panel,
short panel, and landscape panel.
Previously, separate personalization models were trained per canvas, each
relying only on member-asset interaction histories from the corresponding canvas.
This left two big gaps: \emph{title-level} cold-start (newly launched
titles have little interaction data) and \emph{canvas-level} cold-start (signal
from one canvas did not transfer to another).

Pretrained CLIP image embeddings~\cite{radford2021clip} address both
gaps because they encode \emph{visual content} in a representation
shared across image assets, titles, and canvases:

\begin{itemize}
\item \textbf{Cold-start coverage.} A newly launched artwork has little to no
  interaction history but has a CLIP embedding from day zero, so a member's
  taste over visual themes, talent, and color palettes can be applied
  immediately, without waiting for that asset's own interactions to accumulate.
\item \textbf{Cross-title transfer.} A member who has consistently
  interacted with artwork that prominently features a particular talent
  across prior titles transfers signal through the CLIP embedding to
  newly launched titles, prioritizing artwork where that talent features
  prominently over alternatives where they are less central.
\item \textbf{Cross-canvas transfer and model consolidation.} Artworks
  for different canvases of the same scene share similar CLIP
  embeddings regardless of crop, resize, or assigned asset ID.
  Collapsing the five per-canvas models into one unified architecture
  enables image-embedding signal flow across canvases, with
  the largest gains for canvases with less training data
  (\S\ref{sec:artwork-ablation}).
\end{itemize}

Specifically, we register and retrieve the per-asset CLIP embedding
$\mathbf{e}_a \in \mathbb{R}^{768}$ from the Netflix Embedding Store,
concatenate it with the learned ID embedding inside the item tower,
and project the result back to the original asset-representation
dimension:
\begin{equation}
  \mathbf{h}_a = \text{MLP}\!\left([\,\mathbf{e}_{\text{id}}(a) ;
    \mathbf{e}_a\,]\right);
  \label{eq:clip-item}
\end{equation}
the rest of the model architecture remains unchanged.

\paragraph{Engineering challenge: balancing canvases during consolidation.}
Training a single model across all five canvases raises a data-mixing problem
that the per-canvas models never faced. The canvases differ substantially in
impression volume, and their rewards are defined by \emph{different interaction
types}. The member action that counts as a positive on one canvas need not be
the one that counts on another. Pooling the raw data would allow the
highest-volume canvas to dominate training, so the low-data canvases we hope to
improve would benefit the least. Hand-tuning per-canvas weights would just
replace that problem with arbitrary hyperparameters or endless online sweeps to
tune them.
Instead we apply a \emph{reward-based weighting}
that follows Netflix's long-term reward
modeling~\cite{tang2023reward,pan2024longterm}. Each training example is weighted
by the long-term reward score attached to its interaction type, so a sample's
influence reflects how much that interaction is worth for long-term member
satisfaction rather than how frequently it occurs.
\begin{equation}
  w(a_{ti}) = \rho\!\left(e(a_{ti})\right),
  \label{eq:reward-weight}
\end{equation}
where $e(\cdot)$ is the type of the observed positive interaction and $\rho$ is
the long-term reward score assigned to that type.
Since each canvas defines its positives differently, this rebalances the canvas
mixture automatically, with no weight set by hand. A canvas contributes according
to the long-term value of the interactions it drives rather than its impression
volume. Consolidation thus becomes feasible, and the unified model optimizes for
long-term satisfaction rather than for whichever short-term action is most
frequent.

\paragraph{Extension: query-aware artwork.}
Because CLIP's text and image encoders share an embedding space, the same
content embedding supports a second use case for free.
In the Netflix Search Page, we blend the personalization score with a
query--image alignment term:
\begin{equation}
  s(x, a, q) = \alpha \cdot \mathbf{h}_x^\top \mathbf{h}_a
             + (1 - \alpha) \cdot \cos\!\left(\mathbf{e}_q^{\text{txt}},\,
               \mathbf{e}_a^{\text{img}}\right),
  \label{eq:query-aware}
\end{equation}
where $\mathbf{e}_q^{\text{txt}}$ is the CLIP text embedding of the
query and the mixing weight $\alpha \in [0,1]$ is tuned through online
A/B testing. No additional modeling effort is required: the CLIP
embeddings already in the item tower carry the text--image alignment.
We present this as a lightweight, low-cost extension. Since it reuses embeddings already served for the
main artwork model, the only added component is the query-side cosine term and
the scalar $\alpha$.

In an online A/B test on the Netflix Search Page, the query-aware variant
increased playthrough rate on the search canvas by
$\mathbf{+0.36\%}$ ($p<0.05$) relative to personalization-only scoring
($\alpha = 1$ in Eq.~\ref{eq:query-aware}).
This gain comes at almost no cost. There is no new model, no new training data,
and no new embedding, only a cosine term against embeddings already served.
Search supplies an explicit statement of member intent that the homepage lacks.
CLIP's shared text--image space then matches that intent against artwork content, with
no additional learning required. 

\subsection{Ablation: CLIP Image Embedding and Model Unification}
\label{sec:artwork-ablation}

Our solution combines two ideas: \emph{consolidating} the five
per-canvas models into a single unified model, and \emph{augmenting} the
item tower with CLIP image embeddings. To isolate each contribution we
ablate along both axes, comparing three variants against the prior
production system of five canvas-specific models:

\begin{itemize}
\item \textbf{V1 -- Image Embedding only.} The existing five per-canvas
  models, each augmented with CLIP image embeddings in its item tower.
  Tests whether bolting content awareness onto the current architecture
  is sufficient.
\item \textbf{V2 -- Unified Model only.} A single model trained across
  all five canvases using only learned ID embeddings, with no image
  features. Isolates the value of cross-canvas consolidation in the
  absence of pretrained content signal.
\item \textbf{V3 -- Unified Model + Image Embedding.} Both together.
\end{itemize}

\paragraph{Offline IPS per canvas.}
Table~\ref{tab:artwork-ablation-offline} reports per-canvas $\Delta$IPS
for each variant relative to the prior per-canvas baseline.
V1 (image embedding alone) lifts IPS on the data-starved short and landscape
panels while leaving the others flat or slightly down,
confirming that pretrained visual content does help, especially where
asset interaction history is sparse. V2 (unified model alone) shows
that consolidation by itself transfers signal across canvases through
shared user representations, again with the largest gains
concentrated on the canvases that previously had the least training
data. V3 combines both ideas and is the strongest variant overall,
with pronounced lifts on the cold-start--heavy surfaces (short
panel, landscape panel, and horizontal panel) and roughly flat
performance on the more mature canvases (billboard and vertical box).

\begin{table}[htbp]
  \centering
  \caption{Offline $\Delta$IPS by canvas for the three model variants,
    relative to the prior per-canvas baseline on that canvas. Bold marks
    lifts above $1\%$, the threshold at which we treat an offline effect as
    materially large.}
  \label{tab:artwork-ablation-offline}
  \begin{tabular}{lccc}
    \toprule
    Canvas            & V1                & V2                & V3                 \\
    \midrule
    Billboard         & -0.318\%          & -0.021\%          & 0.041\%            \\
    Vertical box      & -0.126\%          & -0.301\%          & -0.090\%           \\
    Short panel       & \textbf{1.065\%}  & \textbf{2.243\%}  & \textbf{5.691\%}   \\
    Horizontal panel  & -0.027\%          & 0.245\%           & \textbf{1.087\%}   \\
    Landscape panel   & 0.421\%           & \textbf{1.654\%}  & \textbf{1.667\%}   \\
    \bottomrule
  \end{tabular}
\end{table}

\paragraph{Online A/B results.}
We A/B-tested all three variants against the per-canvas baseline
across all Netflix device platforms for a minimum of four weeks.
Per-canvas decomposition is not available online because members
are exposed to multiple canvases within a session, so we report
the overall core discovery metric
(Table~\ref{tab:artwork-ablation-online}). The online results reveal
a finding that the offline IPS hints at.
\emph{The two ideas are complementary, and neither alone moves
the production metric meaningfully}. V1 and V2 are flat or
non-significant; only V3 produces a substantial and statistically
significant win ($+0.127\%$, $p<0.005$) on the core discovery
metric, a strong lift we attribute primarily to better cold-start performance.
V3 is now in production. Its real test, though, came from the product change on Netflix UI
that motivated this work, which we describe next.

\begin{table}[htbp]
  \centering
  \caption{Online A/B results on the overall core discovery metric
    for the three artwork variants, relative to the per-canvas
    baseline. Per-canvas decomposition is not available online
    because members are exposed to multiple canvases within a
    session. $^{\ast}$ marks a statistically significant lift;
    n.s.\ denotes non-significant.}
  \label{tab:artwork-ablation-online}
  \begin{tabular}{lcc}
    \toprule
    Variant                                & Discovery metric lift        \\
    \midrule
    V1: Image Embedding only               & $-0.0265\%$ (n.s.)            \\
    V2: Unified Model only                 & $+0.0236\%$ (n.s.)         \\
    V3: Unified Model + Image Embedding    & $\mathbf{+0.127\%}^{\ast}$ ($p<0.005$) \\
    \bottomrule
  \end{tabular}
\end{table}

\paragraph{Why are the two ideas complementary?}
The most instructive part of this result is that neither idea works on its own.
V1 and V2 do not move the core metric, and only their combination V3
moves it. We believe this reflects a mutual dependence between content signal and
model capacity. Content embeddings alone (V1) tell each per-canvas model what an asset looks
like, but a model trained on one canvas has too few examples to learn how to
\emph{use} that information in the cold-start phase, where low-data canvases need
it most.
Consolidation alone (V2) gives the model plenty of data, but only ID-based
data, and a newly launched asset has barely any interaction history to draw on.
V3 has both. The abundant impressions on mature canvases teach the shared item
tower how CLIP embeddings map to member preference, and that mapping applies
immediately to new assets on the data-starved canvases.
The offline table supports this reading: the V3 IPS lift on the short panel
($5.691\%$) exceeds the sum of the V1 and V2 lifts, so the two effects
compound rather than simply add.

\paragraph{The motivating challenge: the Eclipse UI launch.}
The work described in this section was designed with a specific product change
in view. Netflix was preparing to launch a new TV UI, internally called
Eclipse~\cite{nytimes2025netflixhome}, its largest home-screen redesign in a
decade, which would shift the dominant artwork canvas from vertical box to short
panel effectively overnight. This was a cold-start problem in its sharpest form.
The canvas about to
receive the most impressions was the one where the per-canvas baseline had the
least training data, and that baseline could only catch up slowly, as
short-panel interactions accumulated. By then Netflix's most visible surface
would be poorly personalized. Waiting for the data to arrive was not an
acceptable plan. Content awareness and model consolidation were our answer.
Consolidation allows short-panel selection to draw on interaction signal combined from
every canvas, and CLIP image embeddings let the unified model personalize a
short-panel asset it has never seen anyone interact with.

We shipped V3 ahead of the launch and measured it with a holdback online A/B
test that ran for a month during the rollout. A small control group remained on
the prior per-canvas production model while all other members were served by V3
(Table~\ref{tab:artwork-holdback}). V3 absorbed the canvas shift immediately.
The gains were larger than in the steady-state ablation above, with
\textbf{+0.233\%} on the core discovery metric and \textbf{+0.184\%} on
streaming hours. We read the gap between the two experiments as evidence for the
mechanism rather than noise, since a sudden change in which canvas dominates is
exactly what V3 was designed for, and the regime where model consolidation and
content embeddings together pay off most.
The lift is also large in absolute terms.
Extrapolated to Netflix's publicly reported ${\sim}191$ billion streaming hours
in 2025~\cite{netflix2025engagement}, a $+0.184\%$ gain corresponds to roughly
$350$ million additional streaming hours per year.

\begin{table}[htbp]
  \centering
  \caption{Holdback results during the launch of Eclipse, comparing
    V3 in production against a small control group held on the prior
    per-canvas model. $^{\ast}$ marks a statistically significant lift.}
  \label{tab:artwork-holdback}
  \begin{tabular}{lcc}
    \toprule
    Metric                  & V3 vs.\ prior production         \\
    \midrule
    Core discovery metric   & $\mathbf{+0.233\%}^{\ast}$              \\
    Streaming hours         & $\mathbf{+0.184\%}^{\ast}$             \\
    \bottomrule
  \end{tabular}
\end{table}

\section{Video Preview Personalization}
\label{sec:video}

Previews are short video segments (typically 30--90 seconds) that autoplay
as members browse the Netflix homepage.
Each show has multiple candidate previews, and our platform selects the most
relevant preview per member.
Prior models relied exclusively on asset IDs, making them content-blind and
severely cold-start-limited.
A first content-aware iteration, \emph{SeqCLIP}~\cite{chen2023invideosearch},
represented each
preview by uniformly sampling its frames, encoding each frame with
a CLIP-like model~\cite{radford2021clip}, and averaging the resulting frame
embeddings into a single preview-level vector. SeqCLIP captured the
visual content of a video preview but, by construction, missed the audio
and dialogue signals that carry emotional tone and narrative,
as well as the chronological arrangement of events.

\subsection{MediaFM: A Tri-Modal Foundation Model}
\label{sec:mediafm}
To capture the full semantic richness of video preview content, we adopt
\emph{MediaFM}~\cite{saluja2026mediafm}, Netflix's first in-house multimodal
foundation model, pretrained on a large-scale corpus of shots from the Netflix
catalog. MediaFM splits a video into shots and represents each shot by fusing
three embeddings into a single $2304$-dimensional vector: \textbf{visual}
(\emph{SeqCLIP}, a CLIP-style~\cite{radford2021clip} encoder), \textbf{audio}
(wav2vec~2.0~\cite{baevski2020wav2vec}), and \textbf{timed-text}
(\texttt{text-embedding-3-large}~\cite{openai2024embeddings}). A
BERT-style~\cite{devlin2019bert} Transformer then encodes the shot sequence,
trained with a self-supervised Masked Shot Modeling objective in the spirit of
masked language~\cite{devlin2019bert} and video
modeling~\cite{tong2022videomae}. Each shot therefore gets a
\emph{contextualized} embedding that reflects not only its own content but its
place in the surrounding narrative. As the intrinsic evaluation in
Appendix~\ref{app:mediafm} shows, this contextualization matters more than the
extra modalities themselves.
Since MediaFM was previously documented only in a company technology blog,
Appendix~\ref{app:mediafm} describes its architecture, pretraining, and
intrinsic evaluation in detail. Here we focus on its \emph{adoption} for video
preview personalization.

\paragraph{From a shot sequence to one item-tower vector.}
We represent a candidate preview by the contextualized shot embeddings that fall
within it, extracted \emph{in context} from the full title sequence, which
\cite{saluja2026mediafm} reports works substantially better than encoding the
preview's shots in isolation. We pool them with an unweighted mean, giving a
preview-level vector $\mathbf{e}_c^{\text{MFM}} \in \mathbb{R}^{2304}$ that enters
the item tower exactly as the CLIP embedding does in Eq.~\ref{eq:clip-item}.

\paragraph{Offline IPS and Online A/B results.}
We evaluated SeqCLIP and MediaFM embeddings against the ID-only baseline both
through offline experiments and through an online A/B test that ran for five
weeks across all Netflix device platforms.
Table~\ref{tab:video-results} reports both
signals side by side. SeqCLIP improves over the ID-only baseline by
introducing visual content awareness, and MediaFM improves further
by adding audio and timed-text modalities and contextualization.
The ordering MediaFM~$>$~SeqCLIP~$>$~ID-only holds on both offline
IPS and the core member streaming metric online, which is consistent with IPS
serving as a useful pre-launch indicator for this task.
The most pronounced gains were observed on TV~UI surfaces, with
statistically significant improvements in TV interaction and streaming
metrics.
Following the experiment, we productized MediaFM~v2.0 as the default
video preview personalization embedding across all device platforms.
As in the artwork experiments, the streaming lift is small only in relative
terms. At our scale it represents a substantial absolute gain in member
streaming hours.

\begin{table}[htbp]
  \centering
  \caption{Offline IPS lift and online A/B lift on the core member
    streaming metric for video preview personalization, relative to the
    ID-only baseline. Both signals agree on the ordering:
    MediaFM $>$ SeqCLIP $>$ ID-only. $^{\ast}$ marks a
    statistically significant online lift.}
  \label{tab:video-results}
  \begin{tabular}{lcc}
    \toprule
    Model             & Offline IPS lift   & Online streaming lift \\
    \midrule
    + SeqCLIP         & $0.332\%$          & $0.187\%^{\ast}$ ($p<0.02$)  \\
    + MediaFM         & $0.380\%$          & $0.193\%^{\ast}$ ($p<0.02$)  \\
    \bottomrule
  \end{tabular}
\end{table}

\paragraph{When does the extra modality help, and when not?}
The online gap between MediaFM and SeqCLIP is small in aggregate
($0.193\%$ vs.\ $0.187\%$) even though the two differ by two entire
modalities. This is expected rather than disappointing. For many previews the
visual track alone is already highly predictive, so audio and timed text are
redundant and the tri-modal model has little room to improve. 
The extra modalities matter most on harder previews, where images are ambiguous
and the tone, comic or tense, comes through in the dialogue, music or score. 
They also help most on the TV surface, where previews autoplay, which is why the
effect is strongest there.

\section{Selecting Embeddings via a Proxy Task}
\label{sec:proxy}

A practical question runs through both systems: when a new foundation
model, model version (e.g., MediaFM~v1 vs.~v2), or modality combination
becomes available, which embedding should we actually plumb
through offline training and online A/B testing?
Each end-to-end trial is expensive: data engineering, model
retraining, embedding backfill, and weeks of member traffic, so
running one per candidate is infeasible.
We therefore gate the end-to-end experimentation funnel with a cheap
proxy task: \emph{predict which asset candidate would win under an unpersonalized,
uniform serving policy, using only the asset content embedding as input}.
A linear probe trained against this target acts as a
sanity check that ranks embeddings before any expensive downstream
investment.

\subsection{The Unpersonalized Popularity-Based Winner}
\label{sec:proxy-winner}

For each show $t$, let $A_t = \{a_{t1}, \dots, a_{tn_t}\}$ denote the
set of candidate assets (e.g., artworks, or video previews)
eligible for that show.
We estimate which asset would be most popular under a uniform,
non-personalized serving policy using exploration data that
randomizes asset exposure. For each asset $a_{ti}$ we log:
\begin{itemize}
  \item $I_{ti}$: number of impressions under exploration,
  \item $C_{ti}$: number of interaction events (e.g., plays),
  \item $p_{ti}$: propensity that $a_{ti}$ was shown.
\end{itemize}
We apply inverse-propensity weighting to obtain
a debiased popularity score for each asset candidate:
\begin{equation}
  \tilde{r}_{ti}
    = \frac{C_{ti}}{I_{ti}\, p_{ti}}.
  \label{eq:debiased-rate}
\end{equation}
The \emph{unpersonalized popularity-based winner} for title $t$ is then
\begin{equation}
  a_t^\ast
    = \arg\max_{a_{ti} \in A_t} \tilde{r}_{ti},
  \label{eq:winner}
\end{equation}
i.e., the asset that performs best under a uniform serving policy after
exposure bias is removed. We convert this to a simple binary
\textsc{is-winner} label at the (title, asset) level,
$y_{ti} = \mathbf{1}[a_{ti} = a_t^\ast]$, with exactly one positive per
title.

\subsection{Probe Task Formulation}
\label{sec:proxy-probe}

Given an embedding $\mathbf{x}_{ti} \in \mathbb{R}^d$ produced by an
upstream representation model, we train a linear classifier
$f_\theta : \mathbb{R}^d \to [0,1]$ (a linear probe in the sense
of~\cite{alain2016probes}) to predict $y_{ti}$ from
$\mathbf{x}_{ti}$ alone, optimizing standard binary cross-entropy:
\begin{equation}
  \mathcal{L}(\theta) = -\sum_{t, i}
    \big[
      y_{ti} \log f_\theta(\mathbf{x}_{ti})
      + (1 - y_{ti}) \log \big(1 - f_\theta(\mathbf{x}_{ti})\big)
    \big].
  \label{eq:probe-loss}
\end{equation}
Critically, the probe receives only the asset embedding: no
title identifier, no cast metadata, no per-title features. This
isolates a question we actually care about. How much of the
variance in non-personalized popularity is recoverable from the
embedding itself? If an embedding encodes the semantic factors
that drive popularity (talent salience, composition, stylistic
cues), then a simple linear classifier should be able to identify
the winning asset from the embedding alone. If those factors are
absent from the embedding, the probe's accuracy will be no better
than picking a candidate at random.

For evaluation, we compare the probe's top-1 accuracy on held-out
titles against the pick-at-random baseline, measuring how well the
embedding recovers the asset ranking.

\subsection{Why a Popularity Proxy Screens Personalization Embeddings}
\label{sec:proxy-justification}

The probe predicts a global winner, while the deployed models rank per member,
so it is worth being precise about what the gate claims. We do not claim that
popularity prediction measures personalization quality. We claim only that an
embedding unable to separate a title's popularity winner from its alternatives
is unlikely to carry the finer distinctions our two-tower models need, since
both tasks draw on the same content factors and popularity is the easier of the
two. The probe therefore prunes clear losers, and the production choice is
always made by offline IPS and online A/B test metrics.

Two details keep the probe from simply rewarding generic, eye-catching artwork.
The label is defined per title (Eq.~\ref{eq:winner}), so the probe must
discriminate among candidates that already share subject and cast, and it is
computed from inverse-propensity-weighted exploration data
(Eq.~\ref{eq:debiased-rate}), which removes exposure bias. Ultimately the gate
is justified empirically: its ranking has so far agreed with offline IPS and
online lift (\S\ref{sec:proxy-application}).

\subsection{Application as an A/B Test Gate}
\label{sec:proxy-application}

We applied the probe to screen a broad set of candidate video
embeddings, including SeqCLIP, several MediaFM variants and versions, and
external multimodal embeddings, prior to any end-to-end model
integration and A/B testing. The probe ranking pruned the candidate
space before any expensive downstream work. The
two top-ranked candidates, SeqCLIP and the leading MediaFM
variant, were carried through the full offline end-to-end experiment and
the online A/B tests.

\begin{figure}[htbp]
  \centering
  \includegraphics[width=\linewidth]{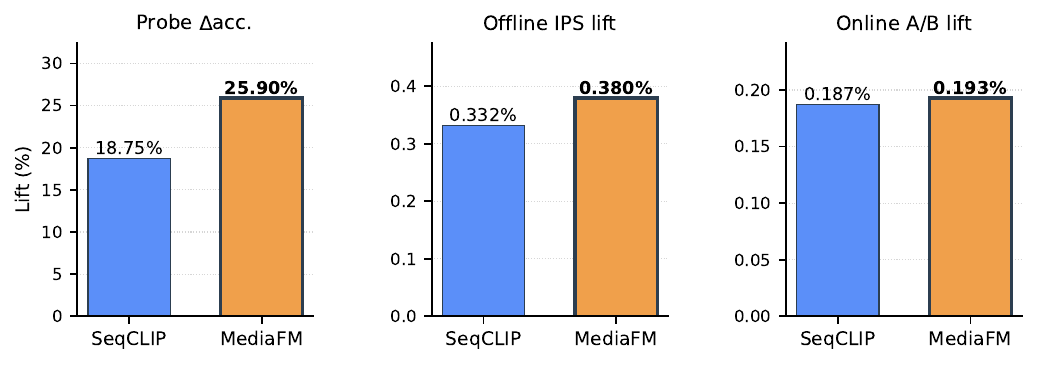}
  \caption{Probe task $\Delta$accuracy against pick-at-random,
    offline IPS lift, and online A/B metric lift for the two
    candidates carried through to full evaluation, relative to the
    ID-based baseline (bold indicates the winner in each pair).
    All three signals agree on the ordering: MediaFM outperforms
    SeqCLIP across the probe, offline IPS, and online A/B.
    Each panel has its own $y$-axis because the three metrics
    live on very different natural scales.}
  \Description{Grouped bar chart with three subplots: probe
    delta-accuracy, offline IPS lift, and online A/B lift. Each
    subplot contains two bars, SeqCLIP and MediaFM, and MediaFM is
    higher in all three: probe delta-accuracy 25.90 vs 18.75 percentage
    points, offline IPS lift 0.380 vs 0.332 percent, online A/B lift
    0.193 vs 0.187 percent.}
  \label{fig:proxy-correlation}
\end{figure}

Figure~\ref{fig:proxy-correlation} compares the three signals side
by side. On held-out titles, MediaFM's top-1 accuracy is $25.90$ percentage points above
pick-at-random and SeqCLIP's is $18.75$ points above. Both are far better than
chance, and MediaFM is the stronger of the two.
The probe task agrees with offline IPS and with the online
A/B metrics. MediaFM is preferred to SeqCLIP across all three, with
the relative ordering preserved across the offline-online boundary.
This is exactly the alignment we rely on when using the probe as a
gate. Candidates that rank below SeqCLIP on the probe have not been
worth the cost of an A/B test in practice. The same proxy probe now
gates every new MediaFM checkpoint before productization.

\section{Conclusion}
\label{sec:conclusion}

We presented two production systems, and one shared method, that use multimodal
embeddings to address cold-start in asset personalization at Netflix. Adding CLIP
image embeddings to a two-tower model makes artwork personalization
content-aware, so a title can be personalized from the day it launches. The same
embeddings, reused in search, let a member's query shape which artwork they see.
For video previews, our tri-modal MediaFM model captures the audio and narrative
signals that visual-only encoders miss, improving both offline IPS and online member metrics.

\subsection{Lessons Learned}
\label{sec:lessons}

Several broader lessons emerged from shipping these systems to production.

\paragraph{1. Consolidating per-canvas models enables CLIP embeddings to carry
signal from high-volume canvases to low-volume ones.}
CLIP embeddings barely change under crop and resize, so the same scene has
nearby representations on every canvas. Folding five per-canvas models into one
therefore lets the abundant impressions on dominant canvases lift the
data-starved ones, which improved both per-canvas offline IPS and the overall
online discovery metric (\S\ref{sec:artwork-ablation}).

\paragraph{2. Content signal and model capacity are complementary, and each
alone may not work properly.}
Content embeddings alone (V1) and consolidation alone (V2) were both flat
online in our A/B test. Only the
combination (V3) won (\S\ref{sec:artwork-ablation}). Before concluding that
content features do not help, check whether a second factor is holding them
back. In our case, the model needed the data that consolidation pooled before it
could learn to use the content features.

\paragraph{3. Multimodality helps most where the visual channel is ambiguous.}
MediaFM beat visual-only SeqCLIP both offline and online, most strongly on TV.
The aggregate gap is modest because the visual track is often enough on its own,
so the extra modalities earn their cost on the harder, dialogue- and
tone-driven part of the catalog (\S\ref{sec:video}). Expect the gain to be
concentrated rather than uniform.

\paragraph{4. A cheap proxy task screens embedding candidates before online test.}
Running an end-to-end A/B test per candidate embedding is prohibitively
expensive once data engineering, retraining, and embedding backfills are counted.
A small linear probe over frozen embeddings (\S\ref{sec:proxy}) prunes the
candidate space before any member traffic is spent, and now gates every new
MediaFM checkpoint.
We use it to rule candidates out, never to pick the
winner (\S\ref{sec:proxy-justification}). It costs almost nothing, and so far it
has not discarded a candidate that would have won.

Looking ahead, we want to use the Embedding Store to place image, text, and
video embeddings in one shared semantic space, in the spirit of
ImageBind~\cite{girdhar2023imagebind}. That would support cross-modal retrieval,
such as matching a video preview to a query or an artwork to a preview, and
unified asset ranking across surfaces.

\begin{acks}
The authors thank Avneesh Saluja, a co-author of MediaFM who no longer at
Netflix, for foundational contributions to the work described here. We also
thank the Netflix Personalization Infrastructure, AI Platform, Evidence Systems,
and Content Engineering teams for their support in developing and deploying
these systems.
\end{acks}

\bibliographystyle{ACM-Reference-Format}
\bibliography{references}

@inproceedings{radford2021clip,
  author    = {Alec Radford and Jong Wook Kim and Chris Hallacy and
               Aditya Ramesh and Gabriel Goh and Sandhini Agarwal and
               Girish Sastry and Amanda Askell and Pamela Mishkin and
               Jack Clark and Gretchen Krueger and Ilya Sutskever},
  title     = {Learning Transferable Visual Models From Natural Language Supervision},
  booktitle = {Proceedings of the 38th International Conference on Machine Learning},
  series    = {Proceedings of Machine Learning Research},
  volume    = {139},
  year      = {2021},
  pages     = {8748--8763},
  publisher = {PMLR},
  address   = {Virtual Event}
}

@inproceedings{xu2016msrvtt,
  author    = {Jun Xu and Tao Mei and Ting Yao and Yong Rui},
  title     = {{MSR-VTT}: A Large Video Description Dataset for Bridging
               Video and Language},
  booktitle = {Proceedings of the IEEE Conference on Computer Vision and
               Pattern Recognition},
  series    = {CVPR '16},
  year      = {2016},
  pages     = {5288--5296},
  publisher = {IEEE},
  address   = {Piscataway, NJ, USA}
}

@article{baevski2020wav2vec,
  author    = {Alexei Baevski and Yuhao Zhou and Abdelrahman Mohamed and
               Michael Auli},
  title     = {wav2vec 2.0: A Framework for Self-Supervised Learning of
               Speech Representations},
  journal   = {Advances in Neural Information Processing Systems},
  volume    = {33},
  pages     = {12449--12460},
  year      = {2020}
}

@inproceedings{covington2016deep,
  author    = {Paul Covington and Jay Adams and Emre Sargin},
  title     = {Deep Neural Networks for {YouTube} Recommendations},
  booktitle = {Proceedings of the 10th ACM Conference on Recommender Systems},
  series    = {RecSys '16},
  year      = {2016},
  pages     = {191--198},
  publisher = {ACM},
  address   = {New York, NY, USA}
}

@inproceedings{huang2013dssm,
  author    = {Po-Sen Huang and Xiaodong He and Jianfeng Gao and Li Deng and
               Alex Acero and Larry Heck},
  title     = {Learning Deep Structured Semantic Models for Web Search
               using Clickthrough Data},
  booktitle = {Proceedings of the 22nd ACM International Conference on
               Information and Knowledge Management},
  series    = {CIKM '13},
  year      = {2013},
  pages     = {2333--2338},
  publisher = {ACM},
  address   = {New York, NY, USA}
}

@article{johnson2021faiss,
  author    = {Jeff Johnson and Matthijs Douze and Herv\'e J\'egou},
  title     = {Billion-Scale Similarity Search with {GPUs}},
  journal   = {IEEE Transactions on Big Data},
  volume    = {7},
  number    = {3},
  pages     = {535--547},
  year      = {2021}
}

@inproceedings{yi2019sampling,
  author    = {Xinyang Yi and Ji Yang and Lichan Hong and
               Derek Zhiyuan Cheng and Lukasz Heldt and Aditee Kumthekar and
               Zhe Zhao and Li Wei and Ed H. Chi},
  title     = {Sampling-Bias-Corrected Neural Modeling for Large Corpus
               Item Recommendations},
  booktitle = {Proceedings of the 13th ACM Conference on Recommender Systems},
  series    = {RecSys '19},
  year      = {2019},
  pages     = {269--277},
  publisher = {ACM},
  address   = {New York, NY, USA}
}

@inproceedings{chandrashekar2017artwork,
  author    = {Ashok Chandrashekar and Fernando Amat and Justin Basilico
               and Tony Jebara},
  title     = {Artwork Personalization at {Netflix}},
  booktitle = {Proceedings of the 11th ACM Conference on Recommender Systems},
  series    = {RecSys '17},
  year      = {2017},
  pages     = {246--250},
  publisher = {ACM},
  address   = {New York, NY, USA}
}

@inproceedings{schnabel2016recommendations,
  author    = {Tobias Schnabel and Adith Swaminathan and Ashudeep Singh and
               Navin Chandak and Thorsten Joachims},
  title     = {Recommendations as Treatments: Debiasing Learning and Evaluation},
  booktitle = {Proceedings of the 33rd International Conference on Machine Learning},
  series    = {Proceedings of Machine Learning Research},
  volume    = {48},
  year      = {2016},
  pages     = {1670--1679},
  publisher = {PMLR},
  address   = {New York, New York, USA}
}

@inproceedings{wei2019mmgcn,
  author    = {Yinwei Wei and Xiang Wang and Liqiang Nie and Xiangnan He
               and Richang Hong and Tat-Seng Chua},
  title     = {{MMGCN}: Multi-modal Graph Convolution Network for
               Personalized Recommendation of Micro-video},
  booktitle = {Proceedings of the 27th ACM International Conference on
               Multimedia},
  series    = {MM '19},
  year      = {2019},
  pages     = {1437--1445},
  publisher = {ACM},
  address   = {New York, NY, USA}
}

@inproceedings{liu2021noninvasive,
  author    = {Sheng Liu and Yifan Hu and Shuai Zhang and Yiqun Liu and
               Min Zhang and Shaoping Ma},
  title     = {Non-invasive Self-attention for Side Information Fusion in
               Sequential Recommendation},
  booktitle = {Proceedings of the 35th AAAI Conference on Artificial Intelligence},
  series    = {AAAI '21},
  volume    = {35},
  year      = {2021},
  pages     = {4249--4256},
  publisher = {AAAI Press},
  address   = {Palo Alto, California, USA}
}

@inproceedings{he2016ups,
  author    = {Ruining He and Julian McAuley},
  title     = {Ups and Downs: Modeling the Visual Evolution of Fashion
               Trends with One-Class Collaborative Filtering},
  booktitle = {Proceedings of the 25th International Conference on World
               Wide Web},
  series    = {WWW '16},
  year      = {2016},
  pages     = {507--517},
  publisher = {International World Wide Web Conferences Steering Committee},
  address   = {Republic and Canton of Geneva, CHE}
}

@misc{chen2023invideosearch,
  author       = {Boris Chen and Ben Klein and Jason Ge and Avneesh Saluja and
                  Guru Tahasildar and Abhishek Soni and Juan Vimberg and
                  Gustavo Carmo and Meenakshi Jindal and Elliot Chow and
                  Amir Ziai and Varun Sekhri and Santiago Castro and
                  Keila Fong and Kelli Griggs and Mallia Sherzai and
                  Robert Mayer and Andy Yao and Vi Iyengar and
                  Jonathan Solorzano-Hamilton and Hossein Taghavi and
                  Ritwik Kumar},
  title        = {Building In-Video Search},
  year         = {2023},
  month        = nov,
  howpublished = {Netflix Technology Blog},
  note         = {\url{https://netflixtechblog.com/building-in-video-search-936766f0017c}}
}

@misc{saluja2023scenechanges,
  author       = {Avneesh Saluja and Andy Yao and Hossein Taghavi},
  title        = {Detecting Scene Changes in Audiovisual Content},
  year         = {2023},
  month        = jun,
  howpublished = {Netflix Technology Blog},
  note         = {\url{https://netflixtechblog.com/detecting-scene-changes-in-audiovisual-content-77a61d3eaad6}}
}

@misc{xu2025qwen3omni,
  author        = {Jin Xu and Zhifang Guo and Hangrui Hu and Yunfei Chu and
                   Xiong Wang and Jinzheng He and Yuxuan Wang and Xian Shi and
                   others},
  title         = {{Qwen3-Omni} Technical Report},
  year          = {2025},
  eprint        = {2509.17765},
  archiveprefix = {arXiv},
  primaryclass  = {cs.CL}
}

@misc{saluja2026mediafm,
  author       = {Avneesh Saluja and Santiago Castro and Bowei Yan and
                  Ashish Rastogi},
  title        = {{MediaFM}: The Multimodal {AI} Foundation for Media
                  Understanding at {Netflix}},
  year         = {2026},
  month        = feb,
  howpublished = {Netflix Technology Blog}
}

@inproceedings{tang2023reward,
  author    = {Gary Tang and Jiangwei Pan and Henry Wang and
               Justin Basilico},
  title     = {Reward Innovation for Long-Term Member Satisfaction},
  booktitle = {Proceedings of the 17th ACM Conference on Recommender Systems},
  series    = {RecSys '23},
  year      = {2023},
  pages     = {396--399},
  publisher = {ACM},
  address   = {New York, NY, USA}
}

@misc{pan2024longterm,
  author       = {Jiangwei Pan and Gary Tang and Henry Wang and
                  Justin Basilico},
  title        = {Recommending for Long-Term Member Satisfaction at
                  {Netflix}},
  year         = {2024},
  month        = aug,
  howpublished = {Netflix Technology Blog},
  note         = {\url{https://netflixtechblog.com/recommending-for-long-term-member-satisfaction-at-netflix-ac15cada49ef}}
}

@misc{openai2024embeddings,
  author       = {{OpenAI}},
  title        = {New Embedding Models and {API} Updates},
  year         = {2024},
  howpublished = {\url{https://openai.com/blog/new-embedding-models-and-api-updates}}
}

@misc{nytimes2025netflixhome,
  author       = {Nicole Sperling},
  title        = {Netflix Unveils a New Home Screen, Its Biggest Redesign in a Decade},
  year         = {2025},
  month        = may,
  howpublished = {The New York Times},
  note         = {\url{https://www.nytimes.com/2025/05/07/business/media/netflix-new-home-screen.html}}
}

@inproceedings{devlin2019bert,
  author    = {Jacob Devlin and Ming-Wei Chang and Kenton Lee and
               Kristina Toutanova},
  title     = {{BERT}: Pre-training of Deep Bidirectional Transformers for
               Language Understanding},
  booktitle = {Proceedings of the 2019 Conference of the North American
               Chapter of the Association for Computational Linguistics:
               Human Language Technologies},
  series    = {NAACL-HLT '19},
  year      = {2019},
  pages     = {4171--4186},
  publisher = {Association for Computational Linguistics},
  address   = {Minneapolis, Minnesota}
}

@inproceedings{tong2022videomae,
  author    = {Zhan Tong and Yibing Song and Jue Wang and Limin Wang},
  title     = {{VideoMAE}: Masked Autoencoders are Data-Efficient Learners
               for Self-Supervised Video Pre-Training},
  booktitle = {Advances in Neural Information Processing Systems},
  series    = {NeurIPS '22},
  volume    = {35},
  year      = {2022},
  pages     = {10078--10093},
  publisher = {Curran Associates, Inc.},
  address   = {Red Hook, NY, USA}
}

@misc{wang2022internvideo,
  author        = {Yi Wang and Kunchang Li and Yizhuo Li and Yinan He and
                   Bingkun Huang and Zhiyu Zhao and Hongjie Zhang and
                   Jilan Xu and Yi Liu and Zun Wang and Sen Xing and
                   Guo Chen and Junting Pan and Jiashuo Yu and Yali Wang and
                   Limin Wang and Yu Qiao},
  title         = {{InternVideo}: General Video Foundation Models via Generative
                   and Discriminative Learning},
  year          = {2022},
  eprint        = {2212.03191},
  archiveprefix = {arXiv},
  primaryclass  = {cs.CV}
}

@inproceedings{girdhar2023imagebind,
  author    = {Rohit Girdhar and Alaaeldin El-Nouby and Zhuang Liu and
               Mannat Singh and Kalyan Vasudev Alwala and Armand Joulin and
               Ishan Misra},
  title     = {{ImageBind}: One Embedding Space to Bind Them All},
  booktitle = {Proceedings of the IEEE/CVF Conference on Computer Vision
               and Pattern Recognition},
  series    = {CVPR '23},
  year      = {2023},
  pages     = {15180--15190},
  publisher = {IEEE},
  address   = {Piscataway, NJ, USA}
}

@article{liu2024mmrecsurvey,
  author    = {Qidong Liu and Jiaxi Hu and Yutian Xiao and Xiangyu Zhao and
               Jingtong Gao and Wanyu Wang and Qing Li and Jiliang Tang},
  title     = {Multimodal Recommender Systems: A Survey},
  journal   = {ACM Computing Surveys},
  volume    = {57},
  number    = {2},
  pages     = {1--17},
  year      = {2024},
  publisher = {ACM},
  address   = {New York, NY, USA}
}

@inproceedings{saito2021counterfactual,
  author    = {Yuta Saito and Thorsten Joachims},
  title     = {Counterfactual Learning and Evaluation for Recommender
               Systems: Foundations, Implementations, and Recent Advances},
  booktitle = {Proceedings of the 15th ACM Conference on Recommender Systems},
  series    = {RecSys '21},
  year      = {2021},
  pages     = {828--830},
  publisher = {ACM},
  address   = {New York, NY, USA}
}

@misc{alain2016probes,
  author        = {Guillaume Alain and Yoshua Bengio},
  title         = {Understanding Intermediate Layers Using Linear Classifier Probes},
  year          = {2016},
  eprint        = {1610.01644},
  archiveprefix = {arXiv},
  primaryclass  = {stat.ML}
}

@misc{netflix2025engagement,
  author       = {{Netflix}},
  title        = {What We Watched: The Netflix Engagement Report
                  ({H1} and {H2} 2025)},
  year         = {2025},
  howpublished = {Netflix},
  note         = {\url{https://about.netflix.com/en/news/what-we-watched-the-first-half-of-2025}
                  and \url{https://about.netflix.com/en/news/what-we-watched-the-second-half-of-2025}.
                  Reports 95B (H1) and 96B (H2) hours viewed, ${\sim}$191B total for 2025.}
}

@inproceedings{song2016thumbnails,
  author    = {Yale Song and Miriam Redi and Jordi Vallmitjana and
               Alejandro Jaimes},
  title     = {To Click or Not To Click: Automatic Selection of Beautiful
               Thumbnails from Videos},
  booktitle = {Proceedings of the 25th ACM International on Conference on
               Information and Knowledge Management},
  series    = {CIKM '16},
  year      = {2016},
  pages     = {659--668},
  publisher = {ACM},
  address   = {New York, NY, USA}
}

\clearpage
\appendix

\section{MediaFM: Architecture, Pretraining, and Intrinsic Evaluation}
\label{app:mediafm}

MediaFM was previously documented only in a Netflix technology
blog~\cite{saluja2026mediafm}. For completeness and to make this paper
self-contained, we give an archival description of the model here. Its
adoption for video preview personalization is covered in \S\ref{sec:video}.
MediaFM serves several Netflix applications. These include ad relevancy, video
clip tagging, and video preview personalization which we discussed in the paper.

\subsection{Motivation and Design Rationale}
\label{app:mediafm-motivation}
Matching members to content requires machine-level comprehension of the entire
catalog, from blockbusters to niche documentaries and, increasingly, formats
such as live events and podcasts. Many of the media-understanding tasks that
matter in this setting are inherently \emph{long-form}: they turn on narrative
dependencies and emotional arcs that unfold across a full episode or film
rather than on the content of any single frame. They are also inherently
\emph{multimodal}. Our earlier work on scene-change
detection~\cite{saluja2023scenechanges} found the audio soundtrack to be a
crucial non-visual signal. Audio helps both to locate where a scene begins and
to characterize the tone of a segment. It is precisely the signal that a
visual-only encoder such as SeqCLIP discards by construction. MediaFM is the
first tri-modal (video, audio, text) model pretrained on the Netflix catalog
itself. Pretraining on the catalog makes its representations
entertainment-specific rather than generic.

MediaFM deliberately emits \emph{embeddings} rather than generated text. A
single frozen representation is produced once and reused across many
downstream services. Per-task generative fine-tuning, by contrast, is
comparatively fragile and has to be repeated for each consumer. The embedding
interface also makes the model extensible: a new modality can be folded into
the fused input without disturbing existing consumers. This is the same
modularity argument that motivates the Embedding Store (\S\ref{sec:overview}).

\subsection{Input Representation}
MediaFM's fundamental unit is a \emph{shot}, a run of consecutive video frames
between camera cuts.
Shots are a sensible granularity, given that they are typically short
($2$--$3$ seconds) and contain few events, objects, and characters.
The shots are computed by segmenting a title (a movie or a TV-show episode)
with a shot-boundary-detection algorithm.
Each shot is encoded in three modalities:
\begin{itemize}
\item \textbf{Visual}: \emph{SeqCLIP}~\cite{chen2023invideosearch}, a
  CLIP-style~\cite{radford2021clip} pretrained model that was subsequently
  fine-tuned on a video-retrieval dataset, embeds frames sampled at uniform
  intervals within the shot.
\item \textbf{Audio}: the shot's audio is embedded with
  wav2vec~2.0~\cite{baevski2020wav2vec}.
\item \textbf{Timed text}: the corresponding timed text (closed captions,
  subtitles, or audio description scripts) is encoded with
  OpenAI's text embedding model \texttt{text-embedding-3-large}~\cite{openai2024embeddings}.
\end{itemize}
The three per-shot embeddings are concatenated and unit-normed into a single
$2304$-dimensional fused shot embedding. A title is represented as a
temporally-ordered sequence of up to $512$ such fused embeddings. Title-level
metadata (e.g., synopsis and tags) is encoded with the same text model. It is
supplied to the sequence through a \textsc{[GLOBAL]} token, so that global
title context is available to every shot. Pretraining uses tens of millions of
individual shots drawn from a large and deliberately diverse slice of the
Netflix catalog. Every pretraining shot carries video and audio. Timed text is
the only routinely missing modality, since shots without dialogue are common.
Shots that lack it are zero-padded in that modality.

\begin{figure}
  \centering
  \includegraphics[width=\linewidth]{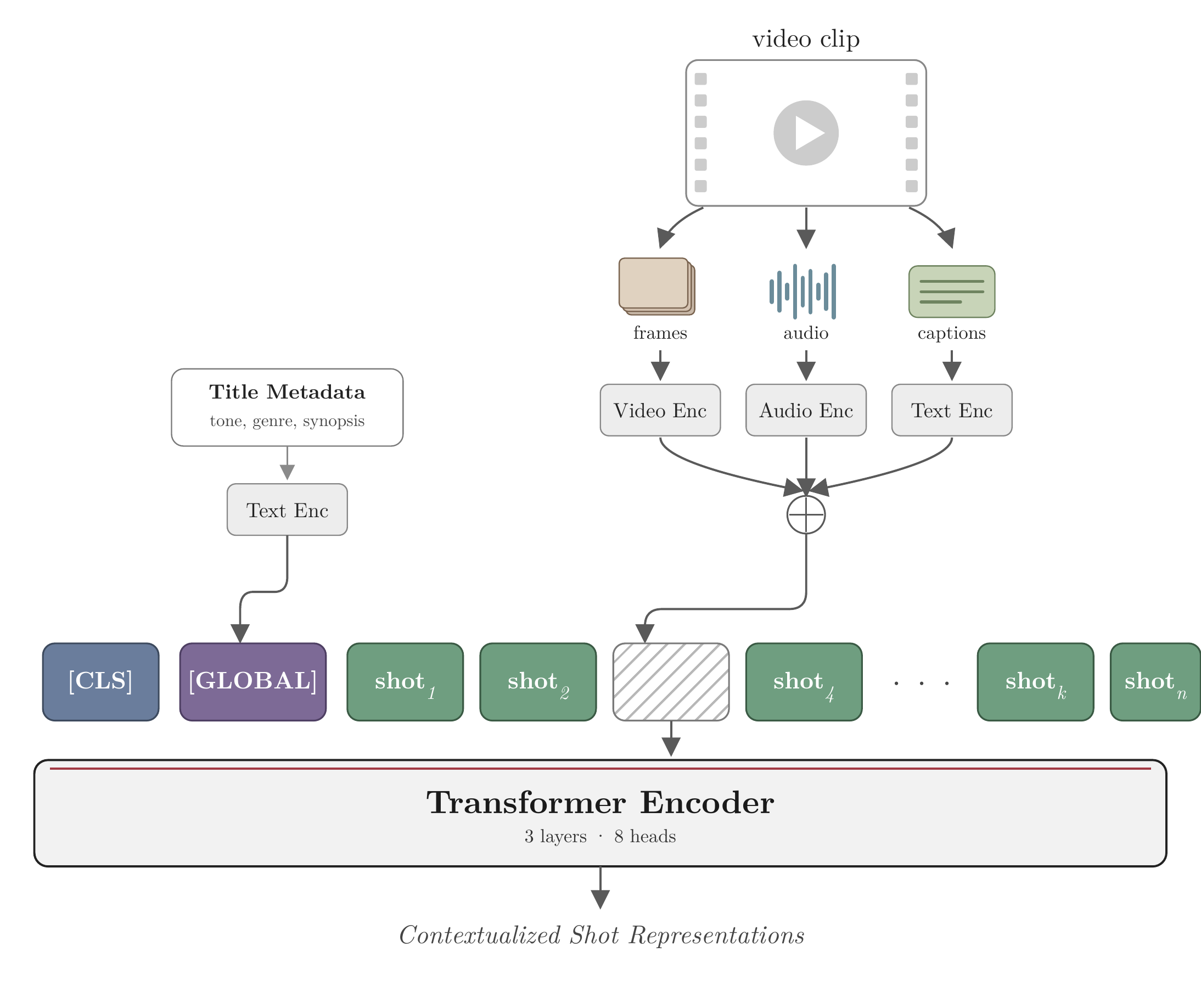}
  \caption{MediaFM architecture. Each shot is encoded in three modalities
    (frames, audio, and timed text) and the three embeddings are fused into a
    single $2304$-dimensional shot vector. Title metadata is encoded with the
    same text model and enters the sequence as a \textsc{[GLOBAL]} token
    alongside a learnable \textsc{[CLS]} token. A three-layer, eight-head
    Transformer encoder then contextualizes the shot sequence. The hatched shot
    is masked under the Masked Shot Modeling objective and must be
    reconstructed by the model. The outputs are contextualized shot
    representations. These are mean-pooled over a preview's shots to form the
    item-tower input (\S\ref{sec:video}).}
  \Description{Block diagram of MediaFM. At the top, a video clip splits into
    frames, audio, and captions, each passed to a video, audio, or text encoder,
    whose outputs are summed into a fused shot embedding. On the left, title
    metadata (tone, genre, synopsis) passes through a text encoder. Both feed a
    token sequence beginning with CLS and GLOBAL tokens followed by shot tokens,
    one of which is hatched to indicate masking. The sequence feeds a Transformer
    encoder with three layers and eight heads, which outputs contextualized shot
    representations.}
  \label{fig:mediafm}
\end{figure}

\subsection{Architecture and Training Objective}
The core of MediaFM is a Transformer encoder architecturally similar to
BERT~\cite{devlin2019bert} (without the next-sentence prediction training
objective). Figure~\ref{fig:mediafm} shows the full architecture.
Processing proceeds in four stages. First, an input projection maps each fused
shot embedding to the model hidden dimension. Second, a learnable
\textsc{[CLS]} token is prepended, and the projected \textsc{[GLOBAL]} metadata
token is inserted immediately after it, so that title-level context
participates in self-attention. Third, positional embeddings are added and the
sequence is contextualized by the Transformer (three layers, eight attention
heads). Fourth, an output projection maps the contextualized states back to the
$2304$-dimensional fused space for prediction.

MediaFM is pretrained with a self-supervised \emph{Masked Shot Modeling} (MSM)
objective, analogous to masked language modeling~\cite{devlin2019bert} and
masked video modeling~\cite{tong2022videomae}. In each sequence, $20\%$ of the
shot embeddings are replaced with a learnable \textsc{[mask]} embedding. The
model then predicts the original fused embedding at each masked position by
minimizing cosine distance to the ground truth. Hidden parameters are optimized
with Muon and the remaining parameters with AdamW. Moving the hidden parameters
off AdamW and onto Muon gave a noticeable improvement in pretraining quality.
Table~\ref{tab:mediafm-config} collects the resulting configuration.

\begin{table}
  \centering
  \small
  \caption{MediaFM configuration.}
  \label{tab:mediafm-config}
  \begin{tabular}{@{}ll@{}}
    \toprule
    Input unit        & Shot (typically 2--3\,s)                      \\
    Visual encoder    & \emph{SeqCLIP}~\cite{chen2023invideosearch}   \\
    Audio encoder     & wav2vec~2.0~\cite{baevski2020wav2vec}         \\
    Text encoder      & \texttt{text-embedding-3-large}               \\
    Fused shot vector & $2304$-d, concatenated, unit-normed           \\
    Sequence length   & Up to $512$ shots per title                   \\
    Special tokens    & \textsc{[cls]}, \textsc{[global]} (metadata)  \\
    Encoder           & BERT-style, 3 layers, 8 heads                 \\
    Objective         & Masked Shot Modeling, $20\%$ masked           \\
    Loss              & Cosine distance to the unmasked vector        \\
    Optimizers        & Muon (hidden), AdamW (rest)                   \\
    \bottomrule
  \end{tabular}
\end{table}

\subsection{Intrinsic Evaluation}
MediaFM is evaluated by training task-specific linear
probes~\cite{alain2016probes} on top of frozen embeddings. Most tasks are
clip-level. For these, we embed a clip \emph{in context}, extracting its shots
from within the surrounding title sequence. This works substantially better
than embedding the clip's shots in isolation. The evaluation suite is:
\begin{itemize}
\item \emph{Clip Popularity Ranking}: predicting a clip's relative
  click-through performance among the other clips of the same title, evaluated
  ten-fold and scored by Kendall's $\tau$.
\item \emph{Clip Tone}: multilabel classification into $100$ tone categories
  (e.g., creepy, scary, humorous) curated by Netflix's internal Metadata \&
  Ratings team, scored by micro-averaged average precision over tones.
\item \emph{Clip Retrieval}: binary ``\emph{clip-worthy}'' classification
  against human annotations, scored by average precision. Each title
  contributes $6$--$10$ positive clips together with matching negatives, at a
  $1{:}3$ positive-to-negative ratio.
\end{itemize}
These probes are decision-support signals rather than end-to-end automation.
Their outputs feed human and team workflows, and the individual consumers are
at varying stages of deployment.

Across all tasks, MediaFM outperforms SeqCLIP as well as external multimodal
embeddings (Google Vertex~AI multimodal embeddings and TwelveLabs
Marengo~2.7). The margins are largest on tasks that require narrative
understanding, such as ad relevancy, where the right ad for a break depends on
the surrounding story context. The external baselines can only be compared on
clip-level tasks. A full title exceeds the maximum video length those APIs
accept, so title-level comparisons against them were not possible.

To separate the contribution of multimodality from that of contextualization,
we compare against a \emph{concat-inputs} baseline. This baseline fuses the
same three per-shot modalities \emph{without} the contextualizing Transformer
(Table~\ref{tab:mediafm-ablation}). Adding modalities without contextualization
yields only modest gains and can even hurt. On clip-popularity ranking, the
concat-inputs baseline underperforms visual-only SeqCLIP, whereas
contextualization consistently drives the largest improvements. Clip retrieval
is the one task where both ingredients pay off evenly. It improves by roughly
$15\%$ relative at each step of the
SeqCLIP~$\rightarrow$~concat-inputs~$\rightarrow$~MediaFM progression. This
finding motivated our adoption of the contextualized MediaFM representation,
rather than a simple multimodal concatenation, for video preview
personalization (\S\ref{sec:video}).

\begin{table}
  \centering
  \small
  \caption{MediaFM intrinsic ablation, isolating contextualization from
    multimodality, as reported in~\cite{saluja2026mediafm}. The
    \emph{concat-inputs} baseline fuses the same three modalities without the
    contextualizing Transformer. Note that on clip-popularity ranking,
    uncontextualized multimodal concatenation \emph{underperforms} visual-only
    SeqCLIP. Higher is better throughout; bold marks the best per row.}
  \label{tab:mediafm-ablation}
  \begin{tabular}{lccc}
    \toprule
    Task (metric)                         & SeqCLIP & Concat-inputs & MediaFM          \\
    \midrule
    Clip Tone (AP)                        & $0.135$ & $0.145$       & $\mathbf{0.166}$ \\
    Clip Popularity Ranking ($\tau$)      & $0.166$ & $0.121$       & $\mathbf{0.207}$ \\
    Clip Retrieval (AP)                   & $0.489$ & $0.561$       & $\mathbf{0.649}$ \\
    \bottomrule
  \end{tabular}
\end{table}

\subsection{Limitations and Next Steps}
\label{app:mediafm-next}
MediaFM's design point is to fuse frozen per-modality encoders at the shot
level, then learn contextualization self-supervised over the catalog. This
keeps pretraining inexpensive and keeps the model extensible to new modalities.
It also has two costs. The model inherits whatever the frozen encoders discard,
and cross-modal fusion is never learned jointly. We are therefore evaluating
pretrained multimodal LLMs in which modality fusion is already learned, such as
Qwen3-Omni~\cite{xu2025qwen3omni}, as a possible base for the next generation
of the model. A parallel line of work extends the same recipe from shot-level
content to title-level metadata embeddings and their adaptation.

\clearpage

\end{document}